# Structural Phase Transition Accompanied by Metal - Insulator Transition in $PrRu_4P_{12}$

C. H. Lee, H. Matsuhata, A. Yamamoto, T. Ohta, H. Takazawa and K. Ueno

*Electrotechnical Laboratory, 1-1-4 Umezono, Tsukuba, Ibaraki 305-8568, Japan*

C. Sekine and I. Shirotani

*Muroran Institute of Technology, 27-1 Mizumoto, Muroran 050-8585, Japan*

T. Hirayama

*JFCC, 2-4-1 Mutsuno, Atsuta-ku, Nagoya 456-8587, Japan*

## Abstract

A structural phase transition has been found using electron diffraction technique in $PrRu_4P_{12}$ accompanied by a metal - insulator (M - I) transition ($T_{MI}$ = 60K). Weak superlattice spots appeared at (H, K, L) (H + K + L = 2n + 1; n is an integer) position at a temperature of T = 12 K and 40 K. Above T = 70 K, the spots completely vanished. The space group of the low temperature phase is probably $Pm\bar{3}$. This is the first observation of a symmetry other than $Im\bar{3}$ in skutterudite compounds.

## 1. Introduction

Filled skutterudite compounds $RM_4X_{12}$ (R = rare - earth; M = Fe, Ru and Os; X = P, As and Sb) show various physical properties, such as magnetic ordering, semiconducting transport property, superconductivity and metal - insulator (M - I) transition. For several years, a great effort has been taken to clarify the origin of the various properties of the skutterudites particularly from the viewpoint of 4f instability. It is expected that further knowledge of the 4f instability can be obtained by studying the physics of the skutterudites. In this paper, we particularly focus our attention on the M - I transition in $PrRu_4P_{12}$.

$PrRu_4P_{12}$ shows a M - I transition at $T_{MI}$ = 60 K [1]. Although several studies have been carried out, the origin of the M - I transition is still a controversial issue. The temperature dependence of the magnetic susceptibility shows no anomaly at $T_{MI}$, suggesting that the M - I transition is not caused by magnetic ordering [1]. Pr $L_2$ - edge XANES measurements indicate that the Pr atoms are almost trivalent independent of temperature [2]. According to powder X - ray diffraction measurements using synchrotron radiation, $PrRu_4P_{12}$ shows no structural phase transition down to T = 25 K [3]. To date, no cause for the M - I transition has been found.

To clarify the mechanism of the M - I transition, determination of the precise crystal structure in the low temperature insulator phase is very important. We expect that there is still a possibility of a small lattice distortion which has been undetected by the powder X - ray diffraction. Thus, we have performed electron diffraction measurements which are very sensitive to lattice distortion, in order to find a structural phase transition at $T_{MI}$.

## 2. Experimental Details

Single crystals of $PrRu_4P_{12}$ were grown by the Sn flux method. High quality Pr (99.9%), Ru (99.9%), P (99.999%) and Sn (99.999%) powders were mixed with the atomic ratio Pr : Ru : P : Sn = 1 : 4 : 20 : 50 and sealed in quartz tubes under a pressure of 150 mm Hg of argon gas. The mixtures were calcined at 1000 °C for one week and then cooled down slowly to 600

°C at 2 °C / h. As - grown samples were treated with aqua regia to dissolve the Sn matrix. The dimensions of the obtained single crystals were typically $0.3 \times 0.3 \times 0.3$ mm$^3$.

The samples were characterized by powder X - ray diffraction at room temperature using Cu K$_{\alpha 1}$ radiation. The crystal structure of PrRu$_4$P$_{12}$ was cubic with a lattice parameter of a = 8.0424 Å, which is consistent with the reported value [4].

For the electron diffraction measurements, the single crystals were mechanically polished and ion etched. Electron diffraction measurements were carried out by using a transmission electron microscope (JEOL 4000FX) operated at 200 kV. A double tilt liquid He holder was equipped for cooling samples down to around 10 K. Specimens could be tilted up to ± 15° along two axes in the microscope.

## 3. Results

Figure 1 shows [0, 0, 1] zone - axis electron diffraction patterns at T = 70 K, 40 K and 12 K. For T = 70 K (Figure 1 (a)), the diffraction pattern is consistent with the Im$\bar{3}$ structure. No spot that was inconsistent with the Im$\bar{3}$ structure was observed at T = 70 K, suggesting that the single crystal was of high quality.

At T = 12 K well below T$_{MI}$ (Figure 1 (c)), weak superlattice spots appeared, for example at the (1, 0, 0) position as indicated by the arrow. This type of superlattice spots was also observed in higher order Laue zones. The positions of the superlattice spots were (H, K, L) (H + K + L = 2n + 1; n is an integer). The clear observation of superlattice spots over a wide **Q** range suggests that this phase transition is accompanied by a lattice distortion. At T = 40 K (Figure 1 (b)), the superlattice spots can still be seen but the intensity becomes very weak, suggesting that a structural phase transition occurs between T = 40 K and 70 K. The precise transition temperature is hard to determine owing to difficulty of quantitative analysis of weak reflections.

## 4. Discussion

Since the intensity of the observed superlattice spots was very weak, the structural distortions in the low temperature phase must be relatively small. Thus, we have assumed that the low temperature phase retains the same point group as the high temperature phase ($m\bar{3}$). Under this assumption, we can show that the crystal structure of the low temperature phase is $Pm\bar{3}$. A minor change in the crystal structure is consistent with the lack of observation of a structural phase transition in the powder X - ray diffraction data. In Figure 2, the crystal structure of $PrRu_4P_{12}$ with the space group $Pm\bar{3}$ is shown. As can be seen in the figure, two sites exist for the Pr and P atoms in the $Pm\bar{3}$ structure, while the $Im\bar{3}$ structure has only one site for each Pr and P atom. In the $Pm\bar{3}$ structure, one possible distortion is the displacement of P (1 or 2) atoms toward or away from the Pr (1 or 2) atoms. An increase or decrease in the bond length of P (1) - P (1) or P (2) - P (2) atoms is also possible.

As shown above, the M - I transition and the structural phase transition occur almost at the same temperature. At present, several models are consistent with the mechanism of the M - I transition taking the structural phase transition into account. Among them, charge ordering with an energy gap is a candidate to explain the M - I transition. According to the XANES measurements, the valence of the Pr atoms remains constant from 300 K down to 20 K [2]. Thus, charge ordering on the Pr sites can be excluded. On the other hand, since two sites also exist for the P atoms in the $Pm\bar{3}$ structure, charge ordering on P sites seems to be more plausible. In this case, the electron rich P site can be expected to shift towards either a Pr or Ru cation.

Another possibility is the opening of a band gap owing to the P displacement. The P displacement makes the volume of Brillouin zone decrease by half and may change the band structure. According to band calculations for $LaFe_4P_{12}$ [5, 6], hole carriers exist mainly on a single Fermi surface. Thus, using the analogy of $LaFe_4P_{12}$, carriers in $PrRu_4P_{12}$ in the $Im\bar{3}$ phase are also expected to be on a single Fermi surface, which suggests that a band gap can easily open by a P displacement. Precise site information is required for further discrimination between possible mechanism.

## 5. Conclusion

We have performed electron diffraction measurements on $PrRu_4P_{12}$ and found weak superlattice spots at (H, K, L) (H + K + L = 2 n + 1) below $T_{MI}$. The crystal structure of the low temperature phase is probably $Pm\bar{3}$. The M - I transition appears to be induced by a structural phase transition.

## Acknowledgements

The authors thank I. Hase, T. Yanagisawa and H. Harima for delightful discussions. This work was supported by the NEDO project with contact number of #98E-a-05-003-2 and funds from the Ministry of International Trade and Industry, Japan.

**Figure Captions**

Figure 1  Electron diffraction patterns of PrRu$_4$P$_{12}$ along the [0, 0, 1] zone-axis at (a) T = 70 K, (b) T = 40 K and (c) T = 12 K. Weak superlattice spots are observed at T = 40 K and 12 K. An arrow depicts one of the superlattice spots.

Figure 2  The crystal structure of PrRu$_4$P$_{12}$ with the space group Pm$\bar{3}$.

# References

[1] C. Sekine, T. Uchiumi, I. Shirotani and T. Yagi, Phys. Rev. Lett. **79**, 3218 (1997).

[2] C. H. Lee, H. Oyanagi, C. Sekine, I. Shirotani and M. Ishii, Phys. Rev. B **60**, 13253 (1999).

[3] C. Sekine et al. (unpublished).

[4] W. Jeitschko and D. Braun, Acta Cryst. **B33**, 3401 (1977).

[5] H. Harima, J. Magn. Magn. Mater. **177-181** (1998) 321

[6] H. Sugawara, Y. Abe, Y. Aoki, H. Sato, M. Hedo, R. Settai, Y. Onuki and H. Harima J. Phys. Soc. Jpn. **69** (2000) 2938

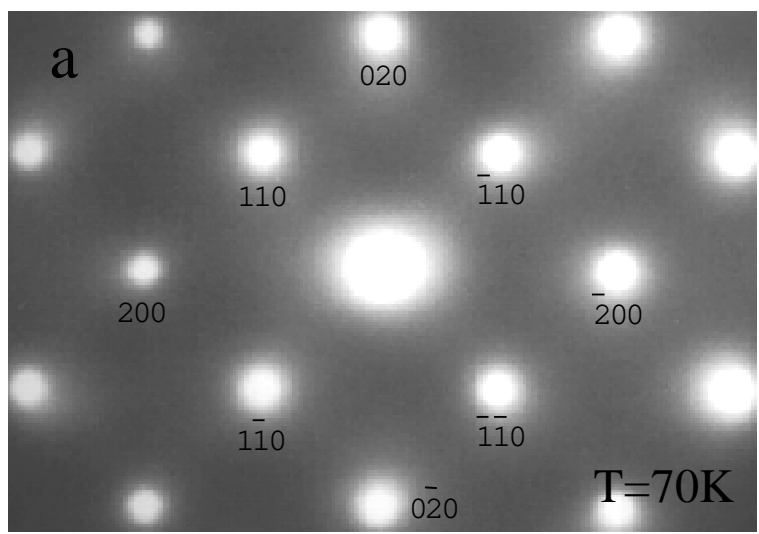

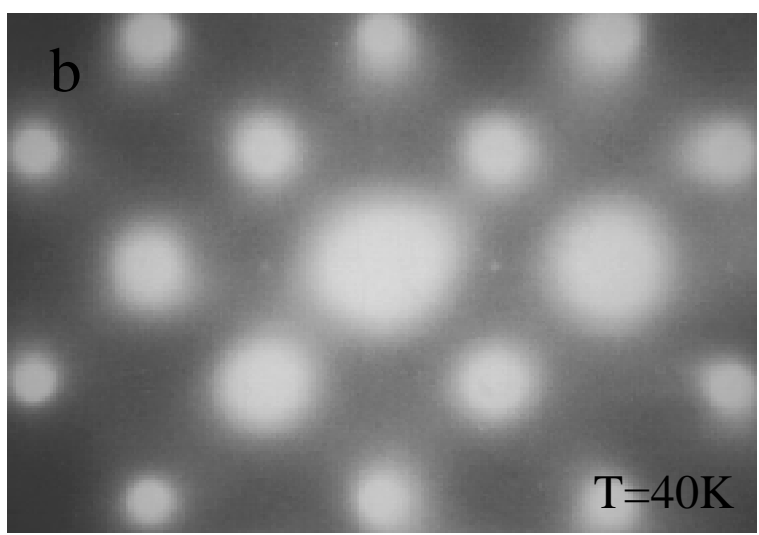

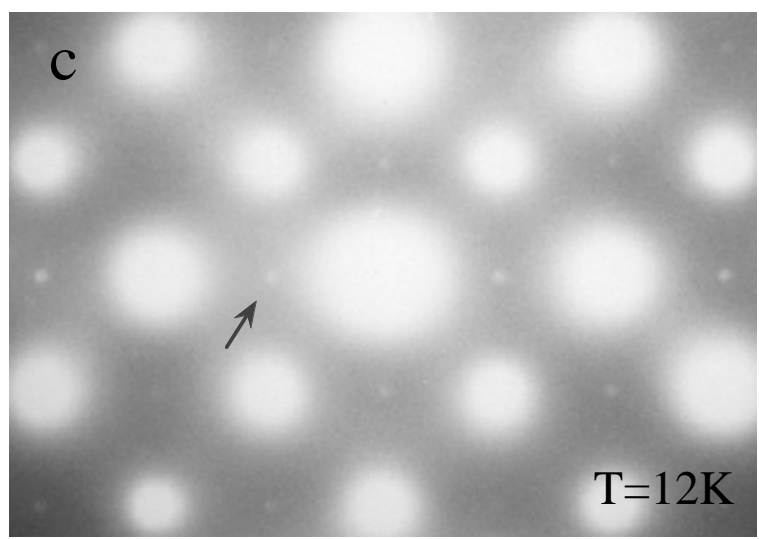

figure 1  C. H. Lee et al.

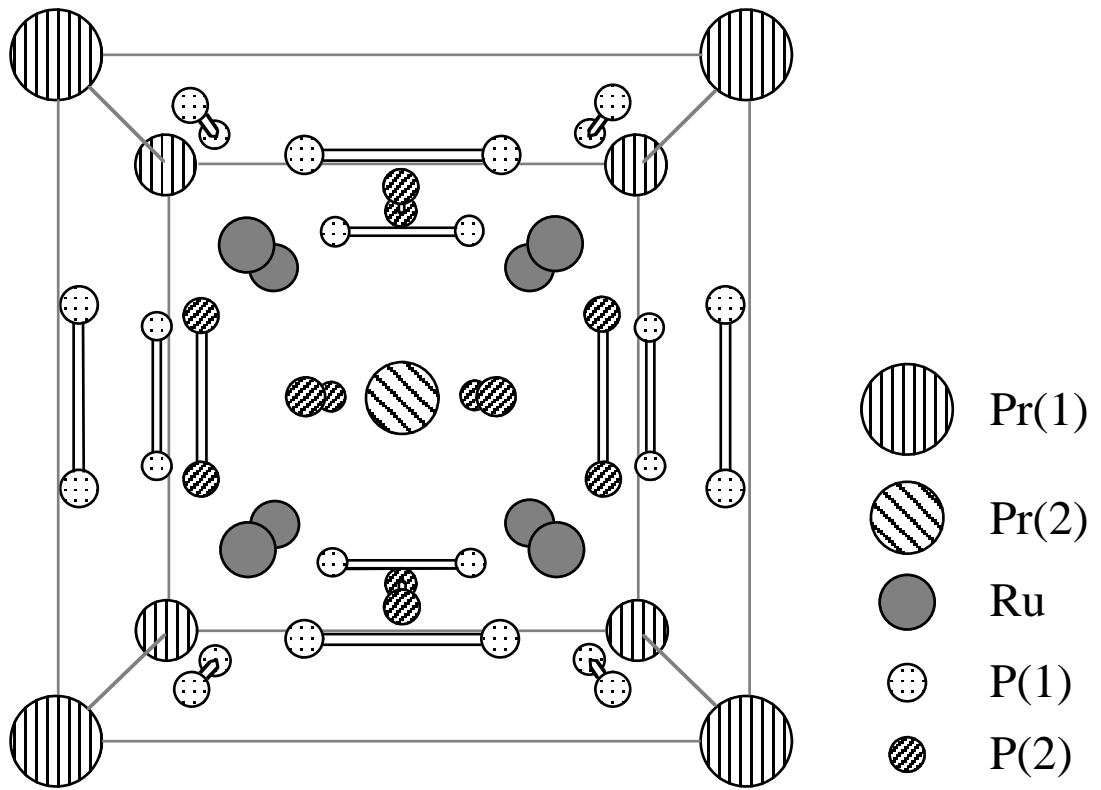